\documentclass[reprint,twocolumn,secnumarabic,amssymb, aps, prl]{revtex4-1}

\usepackage{graphicx}
\usepackage{gensymb}
\usepackage{amsmath}
\usepackage{braket}
\usepackage{lipsum}
\usepackage{appendix}
\usepackage{bm}
\usepackage[colorlinks,linkcolor=black,citecolor=black,urlcolor=black,filecolor=black]{hyperref}
\usepackage{wrapfig}
\begin{document}
\date{\today}

\title{Dispersive optical systems for scalable Raman driving of hyperfine qubits}

\author{H. Levine$^{1}$}
\thanks{Current affiliation: AWS Center for Quantum Computing, Pasadena, CA 91125.}
\author{D. Bluvstein$^{1}$}
\author{A. Keesling$^{1,2}$}
\author{T. T. Wang$^{1}$}
\author{S. Ebadi$^{1}$}
\author{G. Semeghini$^{1}$}
\author{A. Omran$^{1,2}$}
\thanks{Current affiliation: Google Switzerland GmbH.}
\author{M. Greiner$^{1}$}
\author{V. Vuleti\'{c}$^{3}$}
\author{M. D. Lukin$^{1}$}

\affiliation{$^1$Department of Physics, Harvard University, Cambridge, MA 02138, USA  \\ $^2$ QuEra Computing Inc., Boston, MA 02135, USA \\ $^3$ Department of Physics and Research Laboratory of Electronics, Massachusetts Institute of Technology, Cambridge, MA 02139, USA}

\begin{abstract}
Hyperfine atomic states are among the most promising  candidates for qubit encoding in quantum information processing. In atomic systems, hyperfine transitions are typically driven through a two-photon Raman process by a laser field which is amplitude modulated at the hyperfine qubit frequency. Here, we introduce a new method for generating amplitude modulation by phase modulating a laser and reflecting it from a highly dispersive optical element known as a chirped Bragg grating (CBG). This approach is passively stable, offers high efficiency, and is compatible with high-power laser sources, enabling large Rabi frequencies and improved quantum coherence. We benchmark this new approach by globally driving an array of $\sim 300$ neutral $^{87}$Rb atomic qubits  trapped in optical tweezers, and obtain Rabi frequencies of 2~MHz with photon-scattering error rates of $< 2 \times 10^{-4}$ per $\pi$-pulse. This robust approach can be directly integrated with local addressing optics in both neutral atom and trapped ion systems to facilitate high-fidelity single-qubit operations for quantum information processing.
\end{abstract}

\maketitle

\setcounter{secnumdepth}{1}

\section{Introduction}
Trapped neutral atoms and atomic ions are among the most pristine quantum systems for quantum science and engineering. In such systems, quantum bits can be encoded in pairs of atomic levels which are defined in hyperfine ground state manifolds or on narrow optical transitions from a single ground state to a metastable excited state \cite{saffman_quantum_2010,bruzewicz_trapped-ion_2019}. Hyperfine-encoded qubits are particularly attractive due to their transition frequencies in the several gigahertz range, which can be driven either directly with microwave fields, or by two-photon stimulated Raman transitions. While microwaves have been used for high-fidelity control  \cite{brown_single-qubit-gate_2011, sheng_high-fidelity_2018}, Raman transitions offer substantially higher, megahertz-scale Rabi frequencies \cite{yavuz_fast_2006, jones_fast_2007} as well as the opportunity for local addressing of individual qubits separated by micrometer lengthscales.

A variety of experimental approaches have been used to drive stimulated Raman transitions of hyperfine qubits. The conventional approach to Raman driving uses two phase-locked lasers, with a frequency difference equal to the hyperfine splitting \cite{kasevich_atomic_1991, wang_hybrid_2016}. Alternatively, mode-locked optical frequency combs have been used in trapped ion systems, wherein pairs of frequency components combine to drive Raman transitions \cite{hayes_entanglement_2010, campbell_ultrafast_2010, islam_beat_2014, mizrahi_quantum_2014}.
Another approach is based on phase modulation of a single laser to produce low-noise sidebands at the hyperfine frequency \cite{lee_atomic_2003, arias_low_2017}. 
This approach necessitates additional interferometric filtering to suppress destructive interference between sideband pairs, resulting in a loss of usable optical power \cite{lee_atomic_2003}. 
Furthermore, each of these previously demonstrated approaches require active stabilization due to interferometric sensitivity.

In this paper, we demonstrate a new method for Raman driving based on phase modulation followed by reflection from a highly dispersive optical element. The dispersive element, a chirped Bragg grating (CBG), changes the relative phases of the phase-modulated sidebands, converting destructive interference to constructive interference and producing amplitude modulation for driving Raman transitions. We show that the dispersive approach offers high-efficiency conversion from phase modulation to amplitude modulation, enables scaling to high optical power, and is passively stable.

This paper is structured as follows: in part~\ref{sec:AmpModulation}, we review how stimulated Raman transitions induced by a multi-frequency laser field can be understood purely in terms of laser amplitude modulation. In part~\ref{sec:PMtoAM}, we show how dispersive optics can be used to efficiently convert phase modulation to amplitude modulation for driving Raman transitions. In part~\ref{sec:RamanLaserSetup}, we describe our Raman laser system in detail, and in part~\ref{sec:Benchmarking} 
we experimentally benchmark its performance  on an array of $\sim 300$ neutral $^{87}$Rb atomic qubits trapped in optical tweezers.
These results demonstrate that this robust approach to Raman driving enables scalable optical control of hyperfine qubits, with future opportunities to integrate into local optical addressing systems in both neutral atom and trapped ion platforms.

\section{Laser amplitude modulation drives stimulated Raman transitions}
\label{sec:AmpModulation}
Stimulated Raman transitions are two-photon processes which  drive transitions between two atomic ground states $\ket{0}$ and $\ket{1}$ (split by a qubit frequency $\omega_q$) through an intermediate excited state $\ket{2}$. Conventionally, Raman transitions are understood as being driven by a laser field containing two frequency components separated by $\omega_q$, resulting in an effective resonant coupling between the  states $\ket{0}$ and $\ket{1}$ with Raman Rabi frequency $\Omega_\text{eff} = \Omega_0^* \Omega_1 / (2 \Delta)$, where $\Omega_0$ ($\Omega_1$) describes the laser coupling strength from $\ket{0}$ ($\ket{1}$) to $\ket{2}$ and $\Delta$ is the laser detuning from the ground-excited transition frequency (with $\Delta \gg \omega_q)$ \cite{steck_quantum_nodate}. More general laser fields containing many frequency components, such as modulated lasers and mode-locked lasers, drive Raman transitions through all pairs of frequency components in the field which are separated by $\omega_q$ \cite{hayes_entanglement_2010}. For example, a laser field containing many uniformly spaced frequency components according to
$
    \Omega(t) = \Omega_0 \sum_n a_n e^{i n \omega_q t}
    \label{eq:comb_sidebands}
$ results in a Raman Rabi frequency given by \cite{hayes_entanglement_2010}
\begin{equation}
    \Omega_\text{eff} = \frac{|\Omega_0|^2}{2\Delta} \sum_n a_n^* a_{n+1}
    \label{eq:RamanRabiFrequency}
\end{equation}

A useful interpretation 
of eq.~\eqref{eq:RamanRabiFrequency} is that the Raman Rabi frequency is
simply proportional to the amount of laser amplitude modulation at the qubit frequency $\omega_q$, as would be measured on a photodetector.
One can see this by directly computing the laser intensity $|\Omega(t)|^2$, which contains oscillating terms at each frequency multiple of $\omega_q$; the term $\cos(\omega_q t)$ in particular has the same coefficient $\sum_n a_n^* a_{n+1}$ which appears in eq.~\eqref{eq:RamanRabiFrequency}.

\begin{figure}
    \centering
    \includegraphics{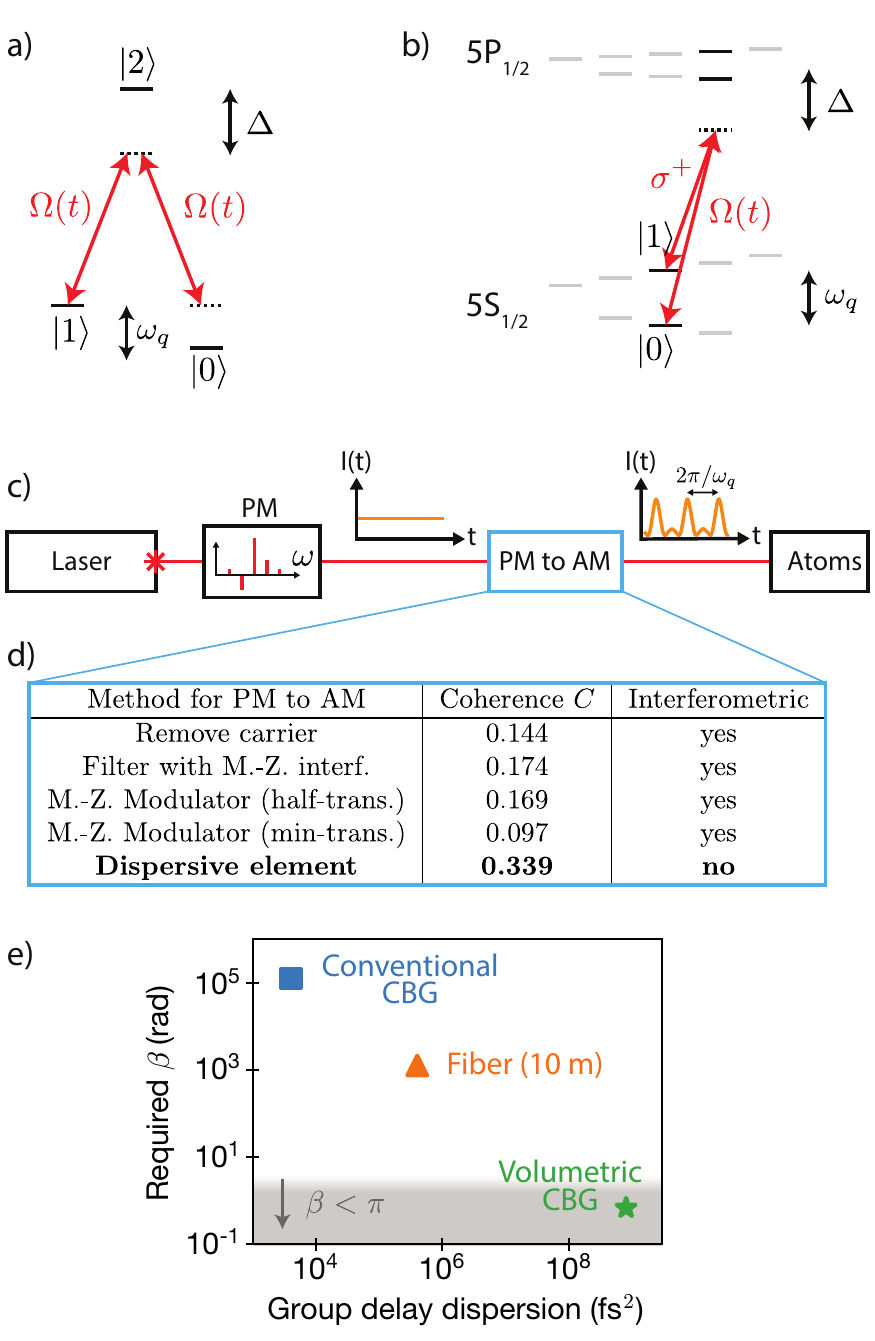}
    \caption{\textbf{Amplitude modulation for driving Raman transitions}. (a) Stimulated Raman transitions in a $\Lambda$-type 3-level system. Adiabatic elimination of the excited state results in an effective Raman coupling between ground states $\ket{0}$ and $\ket{1}$. (b) Level structure for $^{87}$Rb, showing Raman driving of the clock transition from $\ket{0} = \ket{F=1, m_F=0}$ to $\ket{1} = \ket{F=2, m_F=0}$. This transition is driven by a time-dependent $\sigma^+$ polarized field $\Omega(t)$, which is far-detuned by $\Delta$ from the excited state (but not far-detuned relative to the splitting between the $5P_{1/2}$ and $5P_{3/2}$ excited states). (c) Several approaches for Raman driving, including the dispersive approach presented here, operate by converting phase modulation to amplitude modulation at the qubit frequency, which resonantly drives the Raman transition. (d) Comparison of methods for converting phase modulation to amplitude modulation.
    The dispersive approach benefits both from having the highest coherence metric (see Supplement, section II) and from being passively stable since it does not rely on interferometric filtering.
    (e) Weakly dispersive elements, such as a conventional chirped Bragg mirror or a 10~meter optical fiber, require large modulation depth $\beta$ to achieve efficient amplitude modulation, according to eq.~\eqref{eq:dispersion_AM_efficiency}. The highly dispersive volumetric CBG allows a low $\beta$ to be used. $\beta \lesssim \pi$ marks the experimentally accessible window of modulation depths.}
    \label{fig:Figure1}
\end{figure}

This connection between Raman driving and laser amplitude modulation can be further clarified by directly solving the three-level system dynamics in the presence of a generic time-dependent laser field $\Omega(t)$ which couples both ground states $\ket{0}$ and $\ket{1}$ to the excited state $\ket{2}$ (Fig.~\ref{fig:Figure1}a).
This system is described by the following Hamiltonian, given in the rotating frame for the excited state $\ket{2}$:
\begin{align}
\begin{split}
H &= \hbar \omega_{q} \ket{1}\bra{1} + \hbar \Delta \ket{2}\bra{2} \\
&- \frac{\hbar \Omega(t)}{2} \left(\ket{2}\bra{0} + \ket{2}\bra{1} \right) + h.c
\end{split}
\end{align}
If the intermediate detuning $\Delta$ is large compared to $\omega_q$ and the amplitude and spectral width of $\Omega(t)$, we can adiabatically eliminate the excited state, resulting in an effective two-level system (TLS) Hamiltonian for states $\ket{0}$ and $\ket{1}$:
\begin{align}
    H_\text{TLS} &= \hbar \omega_{q} \ket{1}\bra{1}
    - \frac{\hbar \Omega_\text{TLS}(t)}{2} \ket{1}\bra{0} + h.c
    \label{eq:effective_TLS}
\end{align}
with an effective coupling
\begin{equation}
    \Omega_\text{TLS}(t) = \frac{|\Omega(t)|^2}{2 \Delta}
\end{equation}

We highlight here that the Hamiltonian from \eqref{eq:effective_TLS} describes a two-level system with splitting $\omega_{q}$ and time-dependent coupling $\Omega_\text{TLS} \propto |\Omega(t)|^2$. From this description, it is apparent that the \emph{intensity} of the laser field produces an effective field which couples the two qubit states; laser intensity modulation at the qubit frequency therefore drives the qubit transition, akin to resonant driving of a spin transition directly using microwaves. Interestingly, we note that in real atoms (e.g., level structure for $^{87}$Rb as shown in Fig.~\ref{fig:Figure1}b), the ``effective field'' which is proportional to the laser intensity takes the form of the \emph{fictitious magnetic field} associated with vector light shifts (see Supplement, Section I and \cite{cohen-tannoudji_experimental_1972}). Specifically, an off-resonant laser field acts as a fictitious magnetic field given by $\bm{B}^\text{fict} \propto \text{Im} \left[ \bm{\epsilon}^* \times \bm{\epsilon}\right]$, where $\bm{\epsilon}$ is the polarization vector of the laser field \cite{kien_dynamical_2013, thompson_coherence_2013}. Circularly polarized light, such as with $\bm{\epsilon}_+ = \hat{\bm{x}} + i \hat{\bm{y}}$, induces an effective magnetic field oriented along $\hat{\bm{z}}$ which couples $\pi$-polarized spin transitions, and amplitude modulation of the laser field at the transition frequency therefore produces a modulated effective magnetic field which resonantly drives such spin transitions.
This analysis, which extends previous work focusing on vector light shifts in the context of Zeeman transitions \cite{cohen-tannoudji_experimental_1972, mlynek_high-resolution_1981, suter_optically_1990, tanigawa_optical-pumping_1992, deutsch_quantum-state_1998,  zhivun_alkali-vapor_2014}, also clarifies the interplay between laser polarization and Raman transitions.
As an example, the above approach illustrates why linearly polarized light along any propagation axis cannot be used to drive Raman transitions since it produces no vector light shifts, which can be equivalently evaluated through summations over dipole matrix elements.

\section{Efficient conversion of phase modulation to amplitude modulation with dispersive optics}
\label{sec:PMtoAM}

\begin{figure*}
    \centering
    \includegraphics{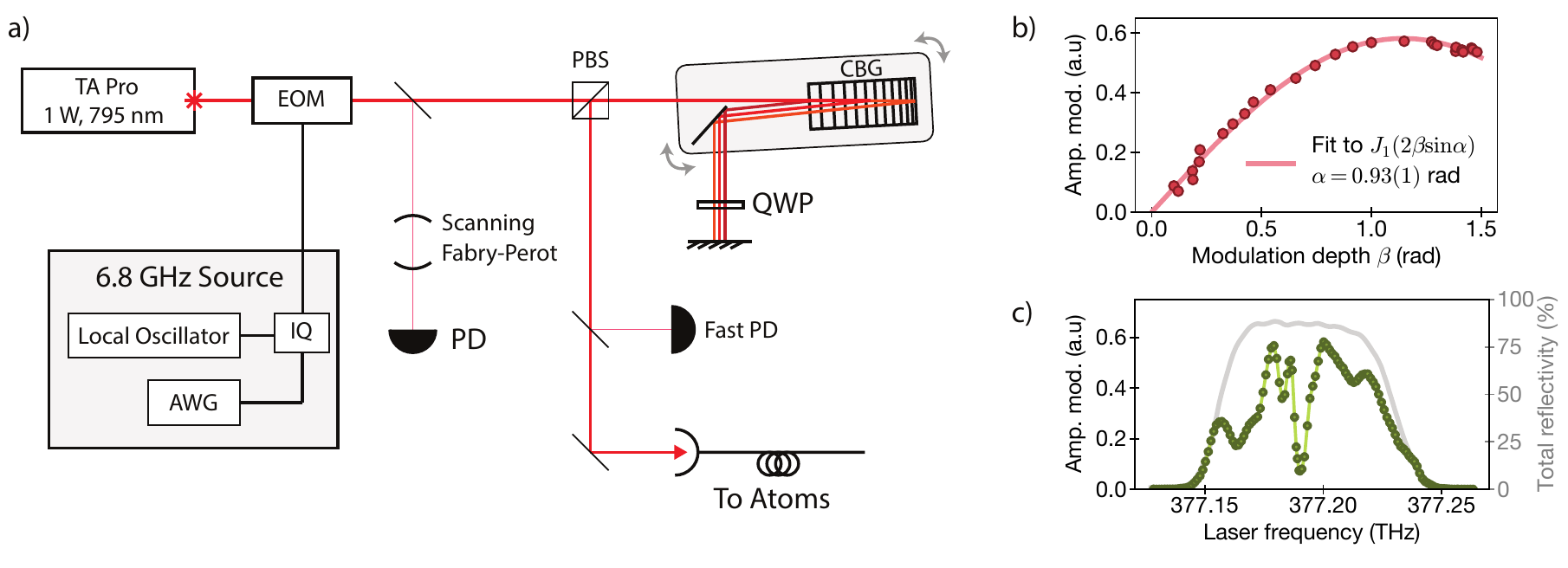}
    \caption{\textbf{Raman laser system using a chirped Bragg grating.} (a) Optical setup. The chirped Bragg grating (CBG) and the first mirror afterwards (in the shaded gray region) are mounted on a single rotation mount. Spectral components separate after the first reflection from the CBG, but recombine after the second reflection. A scanning Fabry-Perot cavity measures the sideband spectrum, and a fast photodetector measures the amplitude modulation. (b) The amplitude modulation depends on both the dispersion of the CBG as well as the phase modulation depth (see main text). We observe the expected Bessel function relation, and can extract the dispersion coefficient. (c) As we scan the laser frequency across the CBG bandwidth, we see a high total reflectivity of the CBG system across the $\sim 50~$GHz  bandwidth. The resulting fiber-coupled light should ideally show constant amplitude modulation across the whole bandwidth, but in practice we observe variation with laser frequency due to nonuniform CBG dispersion over its bandwidth. While more uniform CBGs can be used, the current device can be angle tuned to maximize amplitude modulation, and is insensitive to frequency drifts which are $\lesssim 1$~GHz.}
    \label{fig:Figure2}
\end{figure*}

While laser amplitude modulation is necessary for Raman driving, the most experimentally accessible form of high-frequency laser modulation is \emph{phase modulation} using electro-optics. Sinusoidal phase modulation produces frequency sidebands according to the Jacobi-Anger expansion:
\begin{equation}
    \Omega(t) = \Omega_0 e^{i \beta \sin \omega t} = \Omega_0  \sum_{n=-\infty}^\infty J_n(\beta) e^{i n \omega t}
\end{equation}
where $J_n$ are Bessel functions of the first kind, $\beta$ is the \emph{modulation depth}, and $\omega$ is the modulation frequency. Since the laser intensity is constant $(|\Omega(t)|^2 = |\Omega_0|^2$), a phase-modulated laser cannot  drive hyperfine qubits. This can be seen also as destructive interference between pairs of adjacent sidebands:
$
    \sum_{n=-\infty}^\infty J_n(\beta)^* J_{n+1}(\beta) = 0
$.

There are several methods for modifying the sideband spectrum of a phase-modulated laser to produce amplitude modulation (Fig.~\ref{fig:Figure1}c,d). These methods are primarily interferometric in nature, since they act selectively on frequency components with only gigahertz scale separation. For example, one approach is to use a Fabry-Perot cavity to filter out the carrier ($n=0$) spectral component \cite{yavuz_fast_2006}. Another method is to use a Mach-Zehnder interferometer to filter out all odd-order sidebands, or a Mach-Zehnder intensity modulator in which the phase modulation occurs in one arm of an interferometer \cite{levine_parallel_2019}. These approaches are inherently inefficient, in that they discard some portion of the laser light by filtering out components; further,  they are all sensitive to path-length fluctuations on wavelength scales. Some fiber-based versions of these systems can be more robust, but they are limited to low optical power. Discarding optical power requires detuning the laser system closer to the excited state to achieve the same Rabi frequency, which correspondingly increases the error rate associated with optical scattering \cite{ozeri_errors_2007}. To compare these various approaches, we define a coherence metric $C$ which is proportional to the number of $\pi$-pulses which can be applied before a scattering error (see Supplement, section II). This metric
accounts for how much light is lost in the filtering process as well as how the remaining frequency components interfere, and assumes that the detuning $\Delta$ is chosen to obtain the same Rabi frequency for each approach.
 A high-level comparison of approaches for converting phase modulation to amplitude modulation is presented in  Fig.~\ref{fig:Figure1}d, with details in section II of the Supplement.

\begin{figure*}
    \centering
    \includegraphics{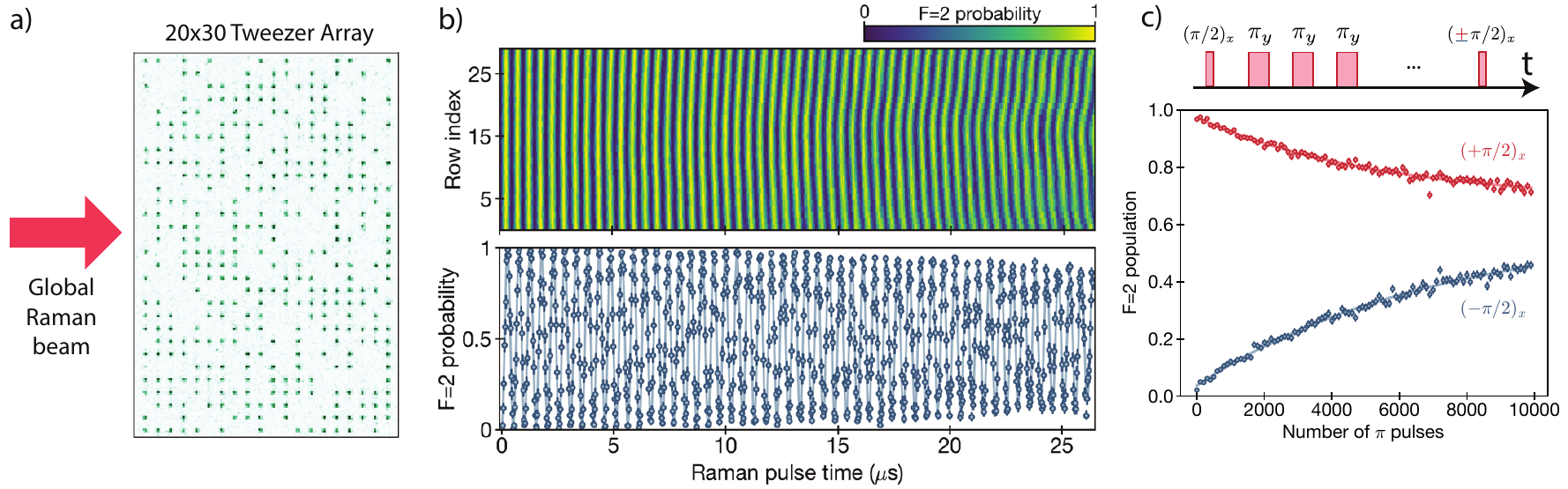}
    \caption{\textbf{Raman driving of $^{87}$Rb atoms in an optical tweezer array.} (a) Sample fluorescence image of $\sim 300$ atoms individually loaded into a 20x30 optical tweezer array. The Raman laser globally illuminates the array. (b) Rabi oscillations, averaged over each row individually (upper panel) and over the middle four rows (lower panel). The measured Rabi frequency is 1.95~MHz. The decay is caused primarily by inhomogeneous averaging across the system. (c) We use a CPMG pulse train to measure how many pulses we can apply before scattering from the Raman laser causes $T_1$-type decay. We compare two measurements in which the final $\pi/2$ pulse is applied along $+x$ (red) or $-x$ (blue), and find that these curves converge with a $1/e$ fit of 7852(76) pulses. This measurement gives a scattering-limited $\pi$ pulse fidelity of 0.999873(1).}
    \label{fig:Figure3}
\end{figure*}

Rather than filtering out specific spectral components from the phase modulation spectrum, we consider here an approach to change the relative phases of these spectral components using dispersive optics. We consider in particular a dispersive element which has a nonzero group-delay dispersion (GDD), defined as
$
    \text{GDD} = \partial^2 \varphi/\partial \omega^2
$.
This element imparts a phase shift to frequency components which is \emph{quadratic} in their frequency; that is, it produces a modified electric field of the form
\begin{equation}
    \Omega(t) = \Omega_0 \sum_{n=-\infty}^\infty J_n(\beta) e^{i n \omega_q t} e^{i\alpha n^2}
\end{equation}
where $\alpha = \text{GDD} \cdot \omega_q^2 / 2$ describes the phase curvature as a function of sideband index. The resulting Raman Rabi frequency depends simply on the phase modulation depth $\beta$ and the dispersion curvature $\alpha$ according to a Bessel function identity (see Supplement, section IV and \cite{zhang_high-frequency_2012}):
\begin{align}
    \Omega_\text{eff} 
    \propto |J_1(2\beta \sin\alpha)|
    \label{eq:dispersion_AM_efficiency}
\end{align}
The Rabi frequency is optimized when the Bessel function $J_1$ is maximized, which occurs when $2\beta \sin \alpha = 1.84$.
However, in practice, electro-optic phase modulation depth is limited to $\beta \lesssim \pi$, requiring $\alpha \gtrsim \pi/4$ to achieve reasonable efficiency; this corresponds to an enormous dispersion of $\text{GDD} \gtrsim 8.5 \times 10^8~$fs$^2$. For comparison, dispersion in a typical optical fiber is $\sim 4 \times 10^4~$fs$^2$/meter
\cite{noauthor_dispersion_nodate, zhang_high-frequency_2012}. Even ultra-high-dispersion chirped Bragg mirrors (mirrors with gradually varying Bragg layer thickness) offer only up to $1300$~fs$^2$ from a single reflection \footnote{Datasheet for Edmund Optics \#12-331, 800nm Highly-Dispersive Ultrafast Mirror} (see Fig.~1e and Supplement, section III for further discussion).


Recently, new optical elements based on \emph{volumetric} Bragg gratings have enabled a new level of frequency selectivity and dispersion control \cite{glebov_volume-chirped_2014}. These crystals have a weak modulation in their refractive index over a lengthscale of $\sim 1$~cm; chirping of the index modulation wavelength as a function of depth produces highly dispersive properties \cite{glebov_volume-chirped_2014}.
We use a chirped volumetric Bragg grating (CBG) with $\text{GDD} = 4\times10^8$~fs$^2$ (OptiGrate, CBG-795-95, apodization of 5~mm on both ends). Reflecting twice from the grating doubles the dispersive effect; this allows us to reach optimal conversion to amplitude modulation with a readily accessible phase modulation depth $\beta \sim 1.3$~rad.
Moreover, the dispersive element does not filter out optical power, but instead produces favorable phase relationships between sidebands, resulting in a high coherence metric (Fig.~\ref{fig:Figure1}d). Finally, the passive stability of the dispersive element simplifies experimental implementation.
Ultimately, the CBG serves as an element which passively converts phase modulation to amplitude modulation, so the effective Raman Rabi frequency (phase, amplitude, and frequency) is directly inherited from the microwave source of the phase modulator.

\section{Raman laser setup}
\label{sec:RamanLaserSetup}

Our Raman laser system (shown in Fig.~\ref{fig:Figure2}a and Supplement, section V) is sourced from a tapered amplifier system which outputs up to $1.5~$W of fiber-coupled optical power at $795$~nm (Toptica TA Pro, free-running at $377.2000$~THz). This light is phase-modulated by a free-space resonant electro-optic modulator (EOM) (Qubig, PM-Rb). The EOM is driven by a 6.8-GHz microwave source, which consists of a frequency-doubled local oscillator (Stanford Research Systems, SG384) that is IQ-modulated by an arbitrary waveform generator (Spectrum Instrumentation, DN2.662-04) to achieve arbitrary frequency, phase, and amplitude control of the phase modulation signal. The laser is then reflected twice from a CBG to convert phase modulation to amplitude modulation, and the output is gated by an acousto-optic modulator (AOM) and coupled into a single-mode fiber. The phase modulation depth $\beta$ is measured by a pickoff onto a scanning Fabry-Perot cavity, and the amplitude modulation is characterized on a fast photodetector (Fig.~\ref{fig:Figure2}a).

The operational bandwidth of the CBG is $50$~GHz; angle tuning of the CBG around the $3^\circ$ target angle of incidence allows shifting of this bandwidth relative to the laser frequency. While the CBG nominally has a uniform dispersion within its bandwidth, we find that in practice the dispersion oscillates within its finite bandwidth (Fig.~\ref{fig:Figure2}c); for this reason, it is helpful to have fine control of the incident angle and to monitor the resulting amplitude modulation while tuning the angle.

In order for the entire optical setup to remain aligned while angle-tuning the CBG, it is important to design the CBG pathway such that the output spatial mode upon the second reflection is independent of the tuning angle. Additionally, the different spectral components of laser light penetrate different depths within the CBG and therefore spatially separate. To recombine these spatial components and ensure overall angle-insensitivity, we use a flat-mirror retroreflector to redirect the spatial components back onto the CBG (Fig.~\ref{fig:Figure2}a). Furthermore, we mount both the CBG and the pickoff mirror which immediately follows on the same rotation stage (with the center of the CBG at the rotation origin), such that the retroreflection condition is met for all tuning angles.
The final retroreflection mirror is aligned once and fixed in place prior to further angle-tuning.
This configuration ensures a single, stable output spatial mode for the light exiting the CBG system 
that is independent of the CBG angle, and therefore maintains subsequent alignment while the angle is tuned.

After optimizing the CBG angle to maximize amplitude modulation (as measured on the fast photodiode), we experimentally measure the dependence of amplitude modulation on the phase modulation depth to confirm the expected Bessel function relationship from eq.~\eqref{eq:dispersion_AM_efficiency} and extract the dispersion coefficient (Fig.~\ref{fig:Figure2}b). Finally, at a fixed modulation depth of $\beta \approx 1.2$~rad, we measure the amplitude modulation and total reflectivity of the double-bounce CBG system as we scan the laser frequency across the bandwidth of the CBG to assess sensitivity to frequency drifts of the laser (Fig.~\ref{fig:Figure2}c); we find that amplitude modulation is stable near optimal points for laser frequency drifts of $\lesssim 1~$GHz.

\section{Benchmarking the Raman laser system on a neutral atom array}
\label{sec:Benchmarking}

We test our high-power Raman laser system on neutral $^{87}$Rb atoms which are loaded within an array of 600 optical tweezers in two dimensions using the platform described in Ref.~\cite{ebadi_quantum_2021} (Fig.~\ref{fig:Figure3}a). The optical tweezers, which are arranged in a 100-$\mu$m $\times$ 200-$\mu$m rectangle, are linearly polarized and have a wavelength of $810~$nm. In each experimental cycle, atoms are loaded and then imaged on an electron-multiplied CCD (EMCCD) camera to detect their positions within the array, and their final states are read out by a second image after pushing out atoms in $F=2$ by cycling photons on the $D_2$ transition  $F=2 \to F'=3$. During loading and imaging, the tweezers have a trap depth of $14$~MHz. During Raman driving, the trap depths are lowered to $5$~MHz and an 8.5-G magnetic field is applied \cite{levine_parallel_2019}.

The Raman laser illuminates the atom plane from the side and is cylindrically focused onto the atoms, resulting in an elliptical beam with waists of $40~\mu$m and $560~\mu$m on the minor  and major axes, respectively, with a total average optical power of $150$~mW on the atoms. The large vertical extent enables homogeneity across the atoms without more advanced beam-shaping techniques.  The laser propagates parallel to the magnetic field and is circularly polarized to drive $\sigma^+$ transitions. The laser frequency is tuned $93$~GHz blue-detuned of the $795$-nm transition to the $5P_{1/2}$ excited state. By tuning the EOM drive frequency, the Raman laser can resonantly drive $\pi$-polarized spin transitions in the ground state hyperfine manifold. We use Raman-assisted optical pumping to prepare atoms in $\ket{0} = \ket{F=1, m_F=0}$ \cite{levine_parallel_2019}. Subsequently, the EOM drive frequency is tuned to the clock resonance, and atoms are coupled from $\ket{0}$ to $\ket{1} = \ket{F=2, m_F=0}$.

We globally drive the qubit array and measure Rabi oscillations across the array with frequency $\Omega_\text{eff} = 1.95$~MHz. We analyze Rabi oscillations individually for each row of the array (Fig.~\ref{fig:Figure3}b, upper panel), as well as averaged over the middle four rows (Fig.~\ref{fig:Figure3}b, lower panel). We attribute the decay of Rabi oscillations primarily to inhomogeneity across the array and small ($\lesssim 1 \%$) power fluctuations.

For Raman operation with hyperfine qubits, there is a fundamental tradeoff between Raman Rabi frequency ($\propto \Omega^2 / 2 \Delta$) and incoherent scattering processes ($\propto \Omega^2 / 4 \Delta^2$). For a given target Rabi frequency, higher optical power enables working at a larger intermediate detuning, increasing the ratio of Rabi frequency to scattering rate (proportional to the coherence metric tabulated in Fig.~\ref{fig:Figure1}d). To evaluate this coherence limitation for our high-power system, we apply a $(\pi/2)_x$ pulse followed by a train of $\pi_y$ pulses (Fig.~\ref{fig:Figure3}c); this so-called CPMG sequence \cite{gullion_new_1990} is robust to pulse miscalibrations that limit our observed Rabi coherence time. By varying the total number of $\pi_y$ pulses, we observe a $T_1$-type decay from scattering, with a characteristic $1/e$ scale of $7852 \pm 76$ pulses. This decay constant sets a lower bound on our scattering-limited $\pi$ pulse fidelity of 0.999873(1).

\begin{figure}
    \centering
    \includegraphics{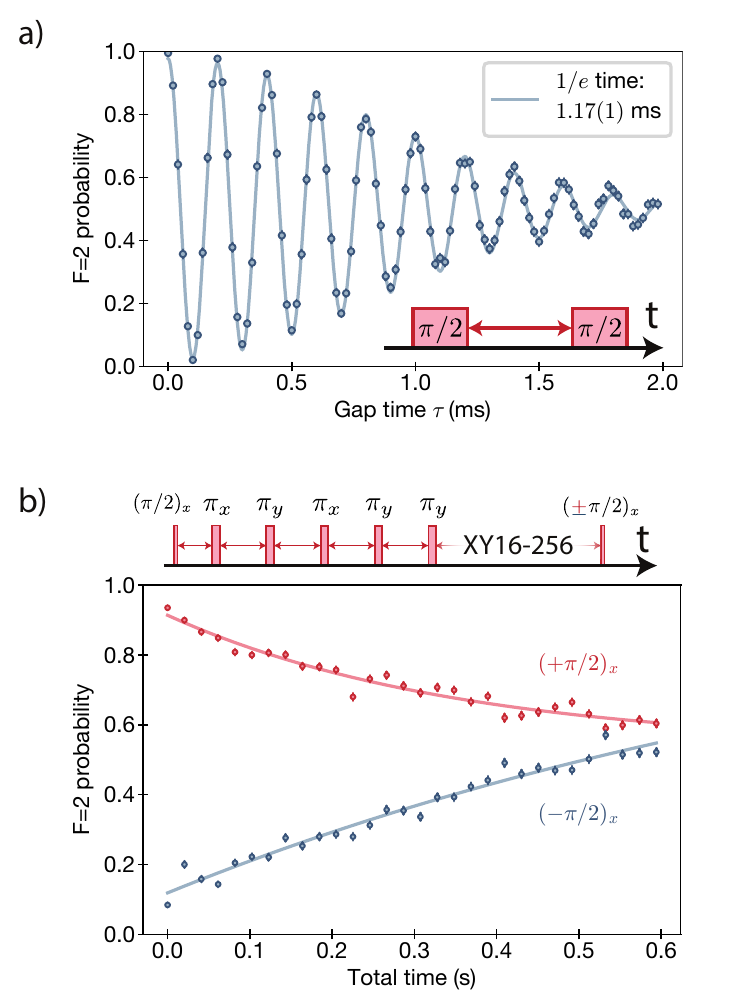}
    \caption{\textbf{Idle coherence of atoms in optical tweezers.} (a) Ramsey measurement, taken with a $5$~kHz detuning between pulses. The atoms occupy several vibrational levels within the tweezers which have different average differential light shifts on the qubit transition, resulting in dephasing. (b) Dynamical decoupling sequence using XY16-256, with a total of 256 $\pi$-pulses. The final $\pi/2$ pulse is applied about $+x$ (red) or $-x$ (blue). These two curves converge with a fitted $T_2 = 303(13)$~ms.}
    \label{fig:Figure4}
\end{figure}

Having established the high Rabi frequency and large number of possible operations in our system, we now explore its utility in preserving coherence across the array, for practical use in quantum information processing protocols. 
We first benchmark the hyperfine coherence in our optical tweezers by measuring a Ramsey $T_2^* = 1.17(1)$~ms (Fig.~\ref{fig:Figure4}a), limited by the finite atomic temperature ($\sim 20~\mu$K) and small differential light shifts in the tweezers ($\sim 4~$kHz) \cite{kuhr_analysis_2005}. By applying a train of $\pi$ pulses, we dynamically decouple the atomic qubits from noise sources such as the tweezer differential light shifts and extend the coherence time to $T_2 = 303(13)$~ms, showing second-timescale coherence across hundreds of qubits (Fig.~\ref{fig:Figure4}b). The $\pi$ pulses are applied according to the XY16-256 pulse sequence (256 total $\pi$ pulses), which is robust against pulse imperfections for generic initial superposition states \cite{souza_robust_2012}. The qubit coherence after the variable-time pulse train is presently limited by residual pulse imperfections, residual dephasing (e.g. fast magnetic field noise or noise on tweezer light shifts), and the $\sim$ 0.5-second $T_1$ time associated with off-resonant scattering from the optical tweezers (see Supplement, section VI). Coherence can be further improved by applying more $\pi$ pulses and by using further-detuned optical tweezers (with trap depth held constant, the tweezer differential lightshifts decrease as $1/\Delta$ and the $T_1$ exhibits a favorable $\Delta^3$ scaling \cite{kuhr_analysis_2005, ozeri_errors_2007}).

Since state-of-the-art Rydberg-based entangling operations are sub-microsecond timescale, and Raman-based single-qubit rotations are also sub-microsecond timescale, the second-scale quantum coherence will allow for a wide variety of deep quantum circuits with hundreds of qubits. Moreover, together with the demonstrated dynamical decoupling sequences, this system should support new approaches for quantum algorithms involving dynamic reconfiguration of atom arrays in sub-millisecond timescales to change the connectivity of Rydberg or photonic cavity-mediated interactions while preserving coherence \cite{ beugnon_two-dimensional_2007, lengwenus_coherent_2010, ordevic_entanglement_2021}.

\section{Conclusion}
While several schemes have been used previously to drive Raman transitions, the dispersive approach offers several advantages. First and foremost, the system is passively stable, and faithfully maps the microwave signal which drives the EOM to the resulting amplitude modulation of the laser field. In contrast, other schemes either require active stabilization of an interferometer, active locking of the repetition rate of a mode-locked laser \cite{hayes_entanglement_2010}, or stabilization of the frequency offset between two combs \cite{islam_beat_2014}. The dispersive approach is additionally more efficient in its use of optical power compared with other approaches using phase modulators. As compared with mode-locked lasers, the experimental simplicity, stability, and low cost make it an attractive alternative.

This dispersive approach can additionally be used for applications in which stimulated Raman transitions are used to couple the atomic spin to motion, such as for Raman sideband cooling or entangling gates in trapped ion systems, akin to the approach taken with mode locked lasers \cite{hayes_entanglement_2010,inlek_quantum_2014}.
Finally, local addressing optics could be used to outcouple the amplitude modulated laser onto individual atoms in the array.
Devices such as spatial light modulators, acousto-optic modulators, and electro-optic modulator arrays can enable fast and parallel control of arbitrary single-qubit rotations in large qubit arrays.
These operations can be integrated with multi-qubit gates based on Rydberg interactions to realize flexible quantum circuits, potentially enabling fully programmable quantum simulations and scalable quantum information processing \cite{saffman_quantum_2010}.

\section{Acknowledgements}
We thank many members of the Harvard AMO community, particularly Elana Urbach, Samantha Dakoulas, and John Doyle for their efforts enabling  operation of our laboratories during 2020-2021. We additionally thank Manuel Endres, Brandon Buscaino, Lev Kendrick, Po Samutpraphoot, and Wes Campbell for helpful discussions.
\textbf{Funding:} We acknowledge financial support from the Center for Ultracold Atoms, the National Science Foundation, the U.S. Department of Energy (DE-SC0021013 \& LBNL QSA Center), the Army Research Office, ARO MURI, and the DARPA ONISQ program. H.L. acknowledges support from the National Defense Science and Engineering Graduate (NDSEG) fellowship. D.B. acknowledges support from the NSF Graduate Research Fellowship Program (grant DGE1745303) and The Fannie and John Hertz Foundation. G.S. acknowledges support from a fellowship from the Max Planck/Harvard Research Center for Quantum Optics.
\textbf{Competing interests:} M.G., V.V., and M.D.L. are co-founders and shareholders of QuEra Computing. A.K. and A.O. are shareholders of QuEra Computing.
\textbf{Data availability:} Data underlying the results presented in this paper are not publicly available at this time but may be obtained from the authors upon reasonable request.

\bibliography{references}
\bibliographystyle{apsrev4-1}

\clearpage
\onecolumngrid
\begin{center}
    \textbf{\large Supplementary Information}\\[.2cm]
    \vspace{0.2cm}
\end{center}

\twocolumngrid

\renewcommand{\thefigure}{S\arabic{figure}}
\renewcommand{\theequation}{S\arabic{equation}}
\renewcommand{\thetable}{S\arabic{table}}
\setcounter{figure}{0}
\setcounter{equation}{0}
\setcounter{section}{0}
\setcounter{secnumdepth}{2}

\section{Driving hyperfine transitions with modulated vector light shifts}

As outlined in the main text, a multi-frequency laser field may be used to drive Raman transitions between hyperfine states if it exhibits amplitude modulation at the hyperfine frequency, $\omega_{HF}$. In this section, we clarify the interpretation of this process through the lens of vector light shifts induced by an off-resonant laser.

We consider an off-resonant laser which couples alkali atoms on the D1 and D2 optical transitions from the $J=1/2$ ground state to the $J'=1/2, J'=3/2$ excited manifolds, respectively. The laser has polarization $\bm{\epsilon}$, field amplitude $\mathcal{E}(t)$ and frequency $\omega(t)$ which is far off-resonance from the D1 and D2 transitions (of frequency $\omega_{D1}$ and $\omega_{D2}$), relative to the hyperfine structure in the excited states.

A traditional analysis of Raman transitions would consider the frequency components of the laser field, encoded in the time dependent amplitude and frequency $\mathcal{E}(t)$ and $\omega(t)$, and would calculate resonant contributions to hyperfine transitions through pairs of components which are separated by $\omega_{HF}$ \cite{SI_hayes_entanglement_2010}. Instead, we will consider the laser field to be slowly varying relative to its large detuning from the excited states, as long as its spectral bandwidth $\Delta \omega$ is small compared to the detuning from the D1 and D2 transitions.

In this regime, the excited states may be adiabatically eliminated and the resulting Hamiltonian for the ground state manifold consists of a scalar light shift (which acts as the identity within the hyperfine manifold, and which we will thus ignore) and a vector light shift term \cite{SI_cohen-tannoudji_experimental_1972, SI_corwin_spin-polarized_1999, SI_kien_dynamical_2013}:
\begin{equation}
    \label{eq:FictitiousFieldHamiltonian}
    H^\text{vec} = \mu_B g_J \bm{B}^\text{fict} \cdot \hat{\bm{J}}
\end{equation}
where $\mu_B$ is the Bohr magneton, $g_J$ is the Land{\'e} factor for the $5S_{1/2}$ levels, and the effective magnetic field is given by
\begin{equation}
    \label{eq:FictitiousField}
    \bm{B}^\text{fict} \propto |\mathcal{E}(t)|^2 \left(\frac{1}{\omega_{D2} - \omega(t)} - \frac{1}{\omega_{D1} - \omega(t)} \right)
    \text{Im}\left[ \bm{\epsilon}^* \times \bm{\epsilon} \right]
\end{equation}

Each term in these expressions offer useful insights into Raman transitions.
Firstly, we note that this effective magnetic field takes the same Hamiltonian form in eq.~\eqref{eq:FictitiousFieldHamiltonian} as a real magnetic field acting on the hyperfine qubit manifold. Just as how a real magnetic field can be modulated using microwave radiation to match a qubit resonance, transitions can be similarly driven within the hyperfine manifold by modulation of $\bm{B}^\text{fict}$ at the hyperfine frequency $\omega_{HF}$ \cite{SI_cohen-tannoudji_experimental_1972, SI_lin_all-optical_2017}.

Secondly, maximizing the effective (Raman) coupling between hyperfine states is achieved by maximizing the modulation amplitude of the fictitious magnetic field in eq.~\eqref{eq:FictitiousField}.
Laser amplitude modulation, consisting of full-scale modulation of $|\mathcal{E}|^2$, is the ideal approach, as described in the main text. Laser phase modulation, which can be understood equivalently as modulation of the frequency $\omega(t)$, plays only a weak role due to the small fractional dependence of $\bm{B}^\text{fict}$ on the laser frequency.

Thirdly, for large detuning from either the D1 or D2 transition, contributions from both states are significant. Tuning the laser frequency $\omega(t)$ in between the two transitions offers constructive interference from both pathways; conversely, detuning far from both states relative to their splitting leads to destructive interference \cite{SI_campbell_ultrafast_2010}.

Finally, the effective magnetic field depends on the laser polarization as $\bm{B}^\text{fict} \propto \text{Im}[\bm{\epsilon}^* \times \bm{\epsilon}]$. While calculating the effects of laser polarization on Raman transitions typically relies on summation over transition matrix elements, the vector light shift interpretation offers a useful alternative.
As a first example, for $\sigma^\pm$ polarized light propagating along the quantization axis, $\bm{\epsilon}_\pm = \hat{\bm{x}} \pm i\hat{\bm{y}}$, resulting in $\bm{B}^\text{fict} \propto \hat{\bm{z}}$; modulation of $\bm{B}^\text{fict}$ along the $\hat{\bm{z}}$ axis therefore drives $\pi$-polarized spin transitions within the ground state manifold. A second example is linearly polarized light, which cannot drive Raman transitions regardless of propagation axis since $\bm{\epsilon}^* \times \bm{\epsilon} = 0$ for linearly polarized $\bm{\epsilon}$. Finally, we consider an example of a circularly polarized laser propagating along $\hat{\bm{x}}$, orthogonally to the quantization axis. For such a laser, with polarization $\bm{\epsilon} = \hat{\bm{y}} \pm i \hat{\bm{z}}$,
the effective field is oriented along $\bm{\epsilon}^* \times \bm{\epsilon} \propto \hat{\bm{x}}$. Just as with a real magnetic field of this orientation, the Raman laser in this configuration couples $\sigma^\pm$ spin transitions within the ground state levels. These examples highlight that interpreting Raman transitions as being driven by modulated vector light shifts offers useful additional intuition beyond the standard analysis of two-photon transitions.

\section{Methods for converting phase modulation to amplitude modulation}
\label{sec:MethodsForPMToAM}

\begin{table*}
    \centering
    \begin{tabular}{|c|c|c|c||c|c|}
    \hline
	Method & Transmission & Amp. Mod. Eff.  & Coherence metric & Optimal phase mod. & \textbf{Max. coherence} \\
    & $T(\beta)$ & $\eta^{AM}(\beta)$ & $C(\beta) = T(\eta^{AM})^2$ & $\beta^*$ (rad) & $C(\beta^*)$\\
	\hline
	Filter out carrier & $1 - J_0(\beta)^2$ & $\frac{2J_0(\beta) J_2(\beta)}{1 - J_0(\beta)^2}$  & $\frac{(2J_0(\beta)J_2(\beta))^2}{1 - J_0(\beta)^2}$ & 3.574 & \textbf{0.144}\\
	Filter with M.-Z. interf. & $\frac{1}{2} (1 + J_0(2\beta))$ & $\frac{J_2(2\beta)}{1 + J_0(2\beta)}$ & $\frac{(J_2(2\beta))^2}{2(1 + J_0(2\beta))}$ & 1.664 & \textbf{0.174}\\
	M.-Z. Modulator (half-transmission) & $1/2$ & $J_1(\beta)$ & $(J_1(\beta))^2/2$ & 1.841 & \textbf{0.169} \\
	M.-Z. Modulator (min-transmission) & $(1 - J_0(\beta))/2$ & $\frac{J_2(\beta)}{1 - J_0(\beta)}$ & $\frac{(J_2(\beta))^2}{2(1 - J_0(\beta))}$ & 2.718 & \textbf{0.097}\\
	\hline
	Dispersive element (coefficient $\alpha$) & 1 & $J_1(2\beta \sin \alpha)$ & $(J_1(2\beta \sin \alpha))^2$ & 1.336 ( $\alpha=0.76$~rad) & \textbf{0.339}\\
	\hline
	Two frequency components & - & 1/2 & - & - & -\\
	$N$ uniform sidebands & - & $\frac{N-1}{N}$ & - & - & - \\
    $N$ optimal sidebands & - & $\cos(\frac{\pi}{N+1})$ & - & - & - \\	
	\hline
  \end{tabular}
    \caption{Comparison of theoretical limits for several approaches for laser amplitude modulation. For approaches based on conversion of phase modulation to amplitude modulation, an overall coherence metric $C$ can be evaluated which is proportional to the number of Rabi oscillations per scattering time, assuming the same total available laser power before filtering. Other approaches, including mode-locked frequency comb lasers, may be compared based on their amplitude modulation efficiency $\eta^{AM}$.}
    \label{tab:Table1}
\end{table*}

\subsection*{Definition of coherence metric}

To evaluate the various methods for converting phase modulation to amplitude modulation, we consider two main parameters for each approach: (1) $T$, the fraction of optical power that is transmitted through the conversion system, and (2) $\eta^{AM}$, the amplitude modulation efficiency of the resulting light. The amplitude modulation efficiency is defined for a field with normalized total power split into uniformly spaced sidebands as $\Omega(t) = \sum_n a_n e^{i n \omega_q t}$, where $\sum_n |a_n|^2 = 1$. In this context, the amplitude modulation efficiency measures how the components interfere to produce amplitude modulation: $\eta^{AM} = \sum_n a_n^* a_{n+1}$. This efficiency is bounded above by 1 and characterizes the Raman Rabi frequency for a fixed total amount of optical power in the system.

The Raman Rabi frequency scales according to $\Omega_\text{eff} \propto T \eta^{AM} / \Delta$, where $\Delta$ is the detuning from the intermediate excited state. At the same time, the rate of optical scattering depends on the average optical power on the atoms, according to $\Gamma_{sc} \propto T / \Delta^2$.

We combine these two parameters into a single metric which best characterizes the coherence properties of each approach. Specifically, we assume a fixed amount of available optical power, and we choose the laser detuning $\Delta$ such that the resulting Raman Rabi frequency $\Omega_\text{eff}$ is fixed. To achieve this, we set $\Delta \propto T \eta^{AM}$. For this setting, the optical scattering scales as $\Gamma_{sc} \propto 1 / T (\eta^{AM})^2$. The ratio of Raman Rabi frequency to scattering rate is therefore given by $\Omega_\text{eff} / \Gamma_{sc} \propto T(\eta^{AM})^2$, which we define as the coherence metric $C$. The comparison of approaches is summarized in Table~\ref{tab:Table1}.


To calculate $T$ and $\eta^{AM}$ for each approach, we begin by considering a phase modulated laser, with (normalized) field:
\begin{equation}
    \Omega(t) = \sum_{n=-\infty}^\infty J_n(\beta) e^{in\omega t}
\end{equation}
The total power is $\sum_n |J_n(\beta)|^2 = 1$. As we evaluate $T$ and $\eta^{AM}$ by considering the filtering of various sidebands, we find that these values can be expressed as simple combinations of Bessel functions through several Bessel function identities (derived in Section~\ref{sec:Bessel_identities}).

\renewcommand\thesubsection{Method~\arabic{subsection}}
\setcounter{secnumdepth}{2}

\subsection{Filter out carrier component}
In this approach, the phase modulation frequency $\omega = \omega_q / 2$, such that frequency components separated by $2\omega$ contribute to the Raman drive of the qubit. After filtering out the carrier, the resulting optical power is
\begin{equation}
    T = 1 - |J_0(\beta)|^2
\end{equation}

The amplitude modulation efficiency is
\begin{align}
    \eta^{AM} &= \left| \left(\sum_{n} J_n(\beta)J_{n+2}(\beta)\right) - J_{0}(\beta) \left(J_{-2}(\beta) + J_2(\beta)\right) \right| / T\\
    &= \frac{2J_0(\beta) J_2(\beta)}{1 - |J_0(\beta)|^2}
\end{align}
The first expression sums up all pairs of frequency components separated with $\Delta n = 2$, and then subtracts the contributions from  $n=0$ with $n=\pm 2$. The sum over all pairs is identically 0, and due to evenness of Bessel functions, $J_{-2}(\beta) = J_2(\beta)$. Complex conjugation in the amplitude modulation efficiency is ignored since the Bessel functions are real-valued.

\subsection{Filter with Mach-Zehnder interferometer}
Here we again consider phase modulation with frequency $\omega = \omega_q / 2$. Passing the laser through a Mach-Zehnder interferometer with a properly chosen path-length difference between arms can result in filtering of all even index or all odd index components in the laser. The optical power after filtering out all odd sidebands (a more favorable configuration) is
\begin{equation}
    T = \sum_{n~\text{even}} J_n(\beta)^2 = \frac{1}{2}(1 + J_0(2\beta))
\end{equation}
due to a Bessel function identity (see Section~\ref{sec:Bessel_identities}). 

The amplitude modulation efficiency in this configuration is also greatly simplified due to a Bessel function identity:
\begin{align}
    \eta^{AM} = \frac{1}{T}\sum_{n~\text{even}} J_n(\beta) J_{n+2}(\beta) &= \frac{1}{T} \left(\frac{1}{2} J_2(2\beta) \right)\\
    &= \frac{J_2(2\beta)}{1 + J_0(2\beta)}
\end{align}

\subsection{Mach-Zehnder modulation}
A Mach-Zehnder modulator is an interferometer in which phase modulation occurs in one arm of the interferometer. If the two pathways are balanced in power, the power transmitted in one output mode is given by the relative phase between the two paths:
\begin{equation}
    I(\phi) = \sin^2(\phi/2) = \frac{1}{2} (1 - \cos(\phi))
    \label{eq:MZModulator}
\end{equation}
To modulate the output intensity at the qubit frequency $\omega_q$, the relative phase can either be biased to the half-transmission point and then modulated at $\omega_q$, according to $\phi = \pi/2 + \beta \sin(\omega_q t)$, or it can be biased to the minimum transmission point and then modulated at $\omega_q/2$, with $\phi = \beta \sin(\omega_q t / 2)$. These approaches result in different electric field components in the output light, but to analyze the Raman performance, we need only analyze the laser intensity.

We begin with the half-transmission configuration. In this case, plugging $\phi = \pi/2 + \beta \sin(\omega_q t)$ into eq.~\eqref{eq:MZModulator}:
\begin{equation}
    I(t) = \frac{1}{2}\left(1 + \sin(\beta \sin (\omega_q t)) \right)
\end{equation}
Using a version of the Jacobi-Anger expansion, the right hand side can be expanded:
\begin{equation}
I(t) = \frac{1}{2} \left(1 - i \sum_{n~\text{odd}} J_n(\beta) e^{i n \omega_q t} \right)
\end{equation}
The average optical power is given by the time-indepenent term:
\begin{equation}
    T = 1/2
\end{equation}
This is as expected, since we modulate symmetrically around the half-transmission point.

The amplitude modulation efficiency is given by the coefficient of the $e^{i \omega_q t}$ term, normalized by $T$:
\begin{equation}
    \eta^{AM} = \frac{1}{T} \frac{J_1(\beta)}{2} = J_1(\beta)
\end{equation}
 
Turning instead to the minimum transmission case, we calculate the time-dependent output intensity by plugging $\phi = \beta \sin(\omega_q t/2)$ into eq.~\eqref{eq:MZModulator}:
\begin{equation}
    I(t) = \frac{1}{2}(1 - \cos(\beta \sin(\omega_q t/2)))
\end{equation}
Again using Jacobi-Anger:
\begin{align}
    I(t) &=
    \frac{1}{2} \left(1 - \sum_{n~\text{even}} J_n(\beta) e^{i n \omega_q t / 2}\right)
\end{align}
We now read off the average optical power by setting all time dependent terms to zero:
\begin{equation}
    T = \frac{1}{2}(1 - J_0(\beta))
\end{equation}
As with the half-transmission case, the amplitude modulation efficiency is the coefficient of the $e^{i \omega_q t}$ term, here corresponding to $n=2$, normalized by $T$:
\begin{equation}
    \eta^{AM} = \frac{1}{T} \frac{J_2(\beta)}{2} = \frac{J_2(\beta)}{1 - J_0(\beta)}
\end{equation}

\subsection{Dispersive elements}
After reflecting from a dispersive element with uniform dispersion (group delay dispersion is independent of frequency), the normalized field is described by:
\begin{equation}
    \Omega(t) = \sum_{n=-\infty}^\infty J_n(\beta) e^{i n \omega t} e^{i \alpha n^2}
\end{equation}
The intensity is then given by:
\begin{equation}
    |\Omega(t)|^2 = \sum_{k=-\infty}^\infty e^{i k \omega t} \sum_{n=-\infty}^\infty  J_n(\beta) J_{n+k}(\beta) e^{i \alpha \left[(n+k)^2 - n^2\right]}
\end{equation}
Assuming the phase modulation frequency is a subharmonic of $\omega_q$, with $\omega = \omega_q / k$, then we have the following amplitude modulation efficiency (of order $k$):
\begin{equation}
    \eta^{AM}_k = \left|\sum_{n=-\infty}^\infty J_n(\beta) J_{n+k}(\beta) e^{2i \alpha n k}\right|
\end{equation}
Here we use the Bessel function identity (see Section~\ref{sec:Bessel_identities}) to simplify:
\begin{equation}
    \eta_k^{AM} = \left|J_k\left(2\beta \sin (\alpha k)\right)\right|
\end{equation}
From this, we can immediately evaluate the upper bound on efficiency for any  choice of $\beta$ and dispersive parameter $\alpha$, because the result is simply bounded by the maximum value of $J_k(z)$. Moreover, we see that modulating directly at $\omega = \omega_q$ (taking $k=1$) is optimal, since $J_1(z)$ has a larger maximum than any higher order Bessel function -- but we also see that this configuration requires the largest dispersive parameter $\alpha$ to achieve this maximum, due to the $\sin(\alpha k)$ coefficient within the Bessel function argument.

\section{Dispersive optical elements}
\label{sec:DispersiveOptics}
The group delay dispersion of an optical element is defined as:
\begin{equation}
    \text{GDD} = \frac{\partial^2 \varphi}{\partial \omega^2}
\end{equation}
where $\varphi(\omega)$ is the optical phase shift (in radians) accumulated by a frequency component with angular frequency $\omega$ after the action of the element. GDD is typically measured in units of fs$^2$, although many optical elements such as fibers have their dispersive properties described in terms of their group velocity dispersion (GVD), which is GDD per unit length (typical units are ps/nm/km).

Normal materials have dispersion which acts over a broad wavelength range, which plays an important role in ultrafast optics with broadband lasers, where dispersion results in pulse broadening. However, we are interested here in strong dispersion on the scale of $\sim 10$~GHz in the near infrared. In particular, as described in the main text, we want optical elements with group delay dispersion of $8 \times 10^8~$fs$^2$ to be able to optimally convert phase modulation to amplitude modulation.

Typical optical fibers at $795~$nm have GVD of $-120$~ps/nm/km, or $4\times 10^4~$fs$^2$/meter, with attenuation $4~$dB/km. To achieve the target GDD, we would require a 20~km fiber, with a resulting 80~dB laser attenuation. Some photonic crystal fibers have been designed to have significantly larger GVD, but with much higher attenuation.

In the ultrafast optics community, after sending short pulses through a long fiber, they reverse the pulse broadening by reflecting the broadened pulse from a chirped Bragg mirror. The highest available chirped Bragg mirrors offer $\text{GDD} \sim 2000~$fs$^2$ per reflection. To achieve our target GDD would require $\sim 400,000$ reflections from such a mirror.

The volumetric chirped Bragg grating (CBG) that we use offers the enormous GDD = $4 \times 10^8$~fs$^2$ from a single pass. After reflecting twice from the CBG, we double the GDD to the target level, and conveniently also recombine spatial modes of all spectral components in the laser. One caveat is that the CBG has a narrow bandwidth of $\sim 50$~GHz, which requires angle-tuning to match to the bandwidth of the phase modulated laser. This could also limit reflectivity at large phase modulation depth due to high order sidebands being outside of the bandwidth, but for $\beta \lesssim \pi$ this does not pose an issue. Another factor is that the CBG does not in fact have uniform GDD over its bandwidth, which further requires angle tuning to position the laser frequency at an optimal point within the CBG bandwidth (Fig.~2c of the main text).

\section{Bessel function identities}
\label{sec:Bessel_identities}

\renewcommand\thesubsection{Identity~\arabic{subsection}}

\subsection{Destructive interference of pure phase modulation}
\label{sec:Bessel_PurePhase}
The Bessel function identities that describe destructive interference in Raman driving with a phase modulated laser can be easily derived from the Jacobi-Anger expansion:
\begin{equation}
    e^{i \beta \sin \omega t} = \sum_{n=-\infty}^\infty J_n(\beta) e^{i n \omega t}
\end{equation}
Taking the magnitude squared of both sides, we find:
\begin{equation}
    1 = \sum_{m,n} J_n(\beta) J_m(\beta) e^{i(m-n)\omega t}
\end{equation}
Regrouping the sum in terms of indices $n$ and $k=m-n$:
\begin{equation}
    1 = \sum_{k=-\infty}^\infty e^{i k \omega t} \left[\sum_{n=-\infty}^\infty J_n(\beta) J_{n+k}(\beta)\right]
\end{equation}
Since the left hand side is time independent, the coefficients of the time dependent terms $e^{i k \omega t}$ must vanish for any $k \ne 0$:
\begin{equation}
    \sum_{n=-\infty}^\infty J_n(\beta) J_{n+k}(\beta) = \begin{cases}
        1 \quad : k = 0 \\
        0 \quad : k \ne 0
    \end{cases}
\end{equation}
Since these sums represent amplitude modulation at frequency $k\omega$, this tautologically says that pure phase modulation has no amplitude modulation.

\subsection{Quadratic phase shifts}
\label{sec:Bessel_quadratic}
\noindent
\textit{Claim:}
\begin{equation}
  J_k(2z \sin \phi) = (-i)^k e^{i k \phi} \sum_{n=-\infty}^\infty  J_n(z) J_{n+k}(z) e^{2i n\phi }
\end{equation}
\textit{Proof}: We begin using the Jacobi-Anger expansion, treating $\beta=2z\sin \phi$ as the modulation depth.
\begin{equation}
  e^{i(2 z \sin \phi)(\sin \theta)} = \sum_{n=-\infty}^\infty J_n(2z \sin \phi) e^{in \theta}
  \label{eq:QuadraticDirectJacobiAnger}
\end{equation}
Alternatively, instead of expanding the left hand side using Jacobi-Anger, we could also multiply the two sine functions, recalling the trigonometric identity:
$
  \sin(x)\sin(y) = \frac{1}{2} \left( \cos(x-y) - \cos(x+y) \right)
$.
Plugging this in, we obtain:
\begin{equation}
  e^{i(2z \sin \phi)(\sin \theta)} = \left( e^{iz \cos(\phi-\theta)}\right) \left(e^{-iz \cos(\phi+\theta)} \right)
\end{equation}
We now apply the Jacobi-Anger expansion for both terms on the right-hand side. Setting this expression equal to the right-hand side of equation~\eqref{eq:QuadraticDirectJacobiAnger}, we obtain:

\begin{multline}
  \left(\sum_{n=-\infty}^\infty i^n J_n(z) e^{i n (\phi-\theta)} \right) \left( \sum_{m=-\infty}^\infty i^m J_m(-z) e^{im(\phi+\theta)} \right) \\
  = \sum_{k=-\infty}^\infty J_k(2z \sin \phi) e^{ik\theta}
\end{multline}
Expanding the left hand side as a sum over indices $n,m$:
\begin{multline}
  \sum_{n,m} i^{n+m} J_n(z) J_m(-z) e^{i(n+m)\phi} e^{i (m-n) \theta} \\ = \sum_{k=-\infty}^\infty J_k(2z \sin \phi) e^{i k \theta}
\end{multline}
We will now rewrite the left hand side with a change in indexing, using $n$ and $k' \equiv m-n$, and regroup terms to pull the $k'$ sum to be the outer sum:
\begin{multline}
   \sum_{k'=-\infty}^\infty e^{ik' \theta} \left[i^{k'} e^{ik' \phi} \sum_{n=-\infty}^\infty i^{2n} J_n(z) J_{n+k'}(-z) e^{2 i n \phi} \right]\\ = \sum_{k=-\infty}^\infty \left[J_k(2z \sin \phi)\right] e^{ik \theta} 
  \label{eq:QuadraticDerivationStep}
\end{multline}
Recalling that $J_{n+k}(-z) = (-1)^{n+k} J_{n+k}(z)$, and using that $i^{2n}=(-1)^n$, we simplify: 
\begin{multline}
  \sum_{k'=-\infty}^\infty e^{ik' \theta} \left[ (-i)^{k'} e^{ik' \phi} \sum_{n=-\infty}^\infty J_n(z) J_{n+k'}(z) e^{2i n \phi}\right] \\= \sum_{k=-\infty}^\infty \left[ J_k(2z \sin \phi) \right] e^{i k \theta}
\end{multline}

In both sides of the equation, we have an outer sum over $k$ (or $k'$), with orthogonal functions $e^{ik\theta}$. We therefore must require that the coefficients are all equal for corresponding $k=k'$. Rewriting the equality between coefficients:
\begin{equation}
  \boxed{
  J_k(2z \sin \phi) = (-i)^k e^{ik\phi} \sum_{n=-\infty}^\infty J_n(z) J_{n+k}(z) e^{2i n \phi}
  }
  \label{eq:QuadraticIdentity}
\end{equation}

\subsection{Even sidebands}
\label{sec:Bessel_evensidebands}
We can now use \eqref{eq:QuadraticIdentity} to prove identities regarding a field with only the even sidebands. We first consider the total power in a beam with only the even-index sidebands:

\textit{Claim:}
\begin{equation}
  T \equiv \sum_{n~\text{even}} J_n(\beta)^2 = \frac{1}{2} \left(1 + J_0(2\beta) \right)
\end{equation}

\textit{Proof:}
We find that the sum over even sidebands is quite similar to a sum over \textit{all} sidebands, but with a minus sign on the odd sidebands. To see this,
\begin{equation}
  \sum_{n=-\infty}^\infty (-1)^n J_n(\beta)^2 = \sum_{n~\text{even}} J_n(\beta)^2 - \sum_{n~\text{odd}} J_n(\beta)^2
  \label{eq:AlternatingSignEvenSidebands}
\end{equation}
Recalling that the sum of the power in all sidebands must be unity, we know that 
\begin{equation}
  \sum_{n~\text{odd}} J_n(\beta)^2 = 1 - \sum_{n~\text{even}} J_n(\beta)^2
\end{equation}
Plugging this into equation \eqref{eq:AlternatingSignEvenSidebands}, we have:
\begin{align}
  \sum_{n=-\infty}^\infty (-1)^n J_n(\beta)^2 &= -1 + 2\sum_{n~\text{even}} J_n(\beta)^2 \\
  &= -1 + 2T
  \label{eq:TotalPowerIdentity}
\end{align}
The left hand side now happens to be in a very similar form to the right hand side of equation \eqref{eq:QuadraticIdentity}. In particular, we now write \eqref{eq:QuadraticIdentity} with $k=0, \phi=\pi/2$, and $z=\beta$:
\begin{equation}
  J_0(2\beta) = \sum_{n=-\infty}^\infty (-1)^n J_n(\beta)^2
\end{equation}
Inserting this result into eq.~\eqref{eq:TotalPowerIdentity}, we solve for $T$:
\begin{equation}
  \boxed{
  T = \frac{1}{2} \left(1 + J_0(2\beta) \right)
  }
\end{equation}

\textit{Claim:} Now we can apply a similar technique to prove another identity related to the situation of even sidebands:
\begin{equation}
  \sum_{n~\text{even}} J_n(\beta) J_{n+2}(\beta) = \frac{1}{2} J_2(2\beta)
\end{equation}

\textit{Proof:}
We begin by directly applying the quadratic dispersion identity \eqref{eq:QuadraticIdentity} with $k=2, \phi=\pi/2$, and $z=\beta$:
\begin{equation}
  J_2(2\beta) = \sum_{n=-\infty}^\infty (-1)^n J_n(\beta) J_{n+2}(\beta)
\end{equation}
Again separating in terms of even and odd terms:
\begin{equation}
  J_2(2\beta) = \sum_{n~\text{even}} J_n(\beta) J_{n+2}(\beta)  - \sum_{n~\text{odd}} J_n(\beta) J_{n+2}(\beta)
\end{equation}

Recalling that the sum over all pairs of sidebands is identically 0, we know that
\begin{equation}
  \sum_{n~\text{odd}} J_n(\beta) J_{n+2}(\beta) = -\sum_{n~\text{even}} J_n(\beta) J_{n+2}(\beta)
\end{equation}

We now plug this result in and find:
\begin{equation}
  \boxed{
  \sum_{n~\text{even}} J_n(\beta) J_{n+2}(\beta) = \frac{1}{2} J_2(2\beta)
  }
\end{equation}

\section{Optical setup}
\label{sec:OpticalSetup}
\begin{figure*}
    \centering
    \includegraphics{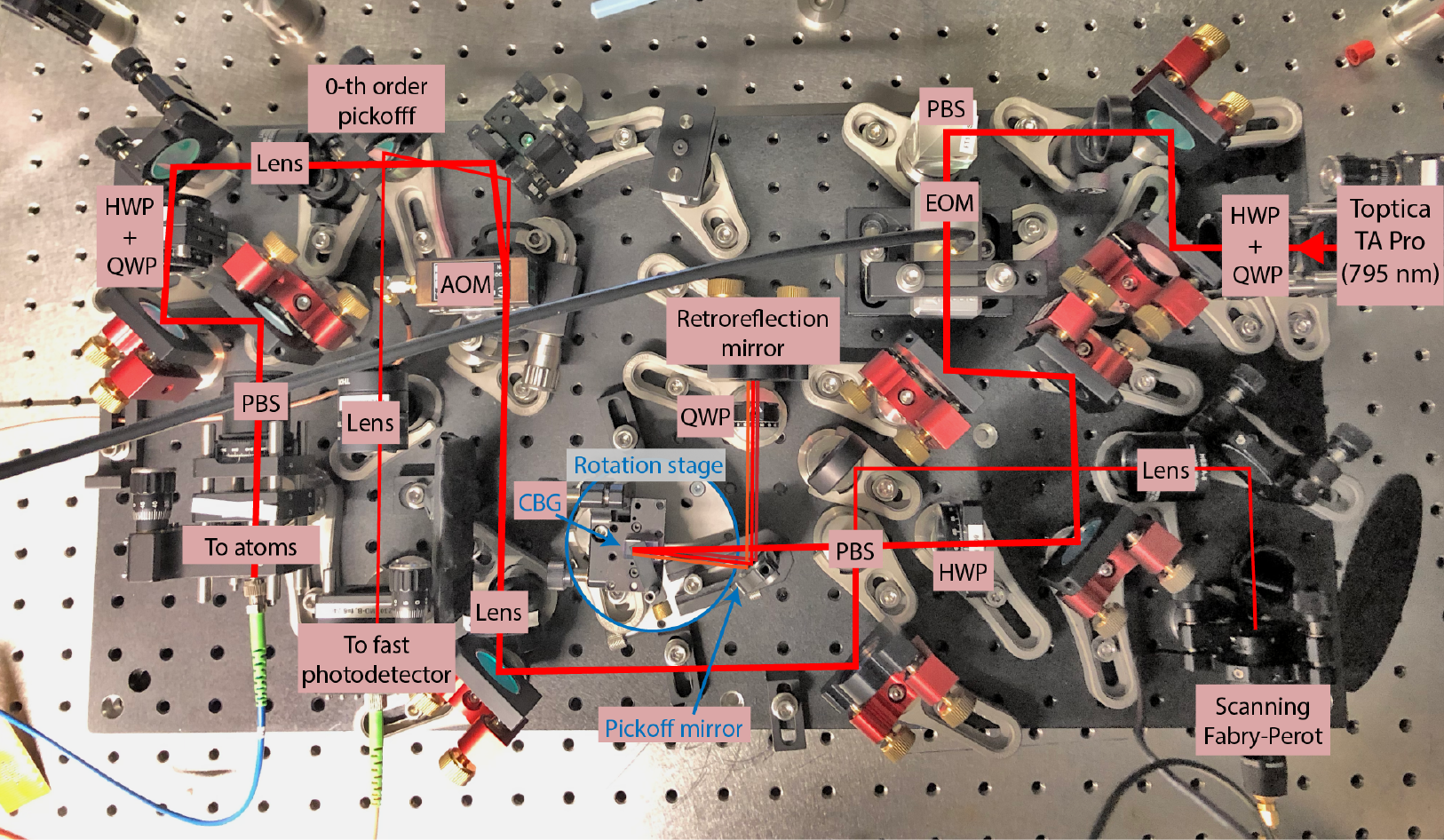}
    \caption{\textbf{Annotated optical setup for the Raman laser system.} TA: tapered amplifier; HWP: half-wave plate; QWP: quarter-wave plate; PBS: polarizing beam splitter; EOM: electro-optic modulator; CBG: chirped Bragg grating; AOM: acousto-optic modulator.}
    \label{fig:OpticalSetup}
\end{figure*}
An annotated image of the optical setup used in this work is shown in Fig.~\ref{fig:OpticalSetup}. The Toptica TA Pro laser source at 795 nm outputs up to 1.5~W of fiber-coupled light. Half-wave and quarter-wave plates align the polarization to be vertical such that it is reflected by a polarizing beamsplitter (PBS) into the electro-optic phase modulator (EOM). A subsequent half-wave plate aligns the polarization to be primarily horizontal such that most of the light propagates through the following PBS (with a small amount deflected up and focused into a scanning Fabry-Perot cavity).

The light transmitted through the PBS reflects from the chirped Bragg grating (CBG) at a $\sim 3^\circ$ angle from normal; a pickoff mirror separates the reflection from the incoming beam. Both the CBG and pickoff are mounted on the same rotation stage, but they have a fixed relative orientation, such that the light is always reflected from the pickoff at a fixed angle upwards. For phase-modulated light, the distinct frequency components penetrate different depths into the CBG and therefore spatially separate; all components reflect from the pickoff mirror at the same angle, however, and are all reflected back onto the CBG by a flat retroreflection mirror. All frequency components pass twice through a quarter-wave plate (QWP) to rotate their polarization such that after recombining on the CBG, they now reflect downwards from the PBS. At this position, the laser is now amplitude modulated. The total reflectivity measured in Fig.~2c of the main text is the ratio of the output power (after the final PBS reflection) to the input power (before the first entrance of the PBS).

Finally, the laser is focused through an acousto-optic modulator (AOM) for power stabilization and fast pulsing. The zeroth-order is aligned into a fiber-coupled fast photodetector for monitoring amplitude modulation. The first-order AOM deflection is coupled into a polarization-maintaining optical fiber and delivered to a separate optical table, where it is outcoupled onto the atoms.

\section{Idle population decay and atom loss in optical tweezers}
\begin{figure*}
    \centering
    \includegraphics{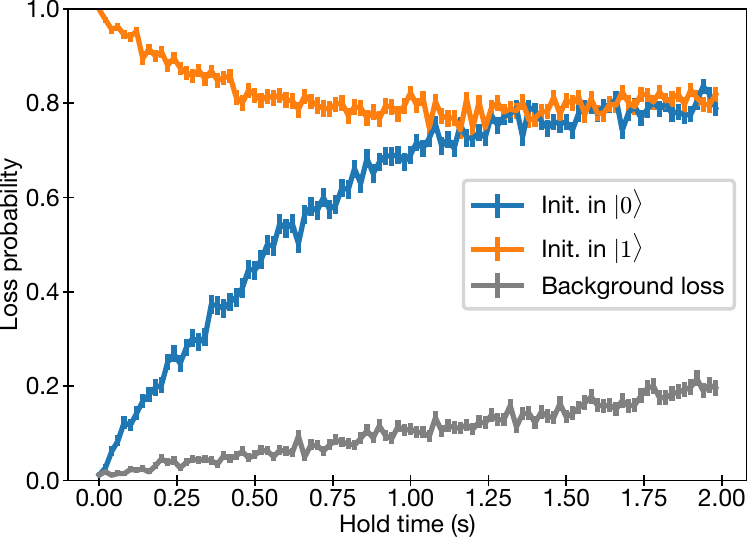}
    \caption{\textbf{Population decay and atom loss.} We initialize atoms in either $\ket{0} = \ket{F=1, m_F=0}$ or $\ket{1} = \ket{F=2, m_F=0}$ (blue and orange curves, respectively), and hold the atoms in the optical tweezers for a variable time before pushing out the $F=2$ population. These two curves converge with a fitted $1/e$ time of $0.45(1)~$s. The tweezer depths are ramped down to $\sim 4~$MHz during the hold time. We additionally turn off the $F=2$ pushout to measure the background loss probability (gray), which is consistent with a 10~second vacuum-limited lifetime.}
    \label{fig:T1_Measurement}
\end{figure*}

While qubit dephasing can be mitigated through dynamical decoupling sequences, the ultimate limit to qubit coherence is set by population decay due to scattering from the optical tweezers, as well as the finite atom lifetime in the tweezers. In Figure~\ref{fig:T1_Measurement}, we show additional data for these two effects, measuring a qubit state population-decay lifetime of $0.45(1)$~s and a background atom lifetime which is $\sim 10~$seconds.

\end{document}